\documentclass[english,aps,preprintnumbers]{revtex4}
\usepackage[T1]{fontenc}
\usepackage[latin9]{inputenc}
\usepackage{amsmath}
\usepackage{amssymb}
\usepackage{graphicx}
\usepackage{babel}
\usepackage[pdftex]{hyperref} 
 
\newcommand*{\eweakgroup}{\mbox{$SU(2)_L \times U(1)_Y$} }
\newcommand*{\emgroup}{\mbox{$U(1)_{em}$} }

\begin{document}

\preprint{BI-TP 2012/08}

\title{Minimizing Higgs Potentials via Numerical Polynomial Homotopy Continuation}

\author{Markos Maniatis}
\email{maniatis@physik.uni-bielefeld.de}
\affiliation{Faculty of Physics, Bielefeld University, 33615 Bielefeld, Germany. }

\author{Dhagash Mehta}
\email{dbmehta@syr.edu}
\affiliation{Physics Department, Syracuse University, Syracuse, NY 13244, USA. }

\begin{abstract}
The study of models with extended Higgs sectors requires to minimize the corresponding
Higgs potentials, which is in general very difficult.
Here, we apply a recently developed method, called numerical polynomial homotopy
continuation (NPHC), which guarantees to find \textit{all} the stationary
points of the Higgs potentials with polynomial-like nonlinearity. 
The detection of all stationary points reveals the structure
of the potential with maxima, metastable minima, saddle points besides
the global minimum.
We apply the NPHC method to the most general Higgs potential having two
complex Higgs-boson doublets and up to five real Higgs-boson singlets.
Moreover the method is applicable to even more involved potentials.
Hence the NPHC method allows to go far beyond the limits
of the Gr\"obner basis approach.
\end{abstract}

\maketitle

\section{Introduction}
\label{s:introduction}
The Standard Model~(SM) comes with a simple Higgs potential:
due to gauge invariance and renormalizability, there is only one Higgs doublet
which appears in the potential in one bilinear and one quartic term.
In particular, finding the global minimum of the Higgs potential,
which gives the vacuum-expectation value of the Higgs field at the stable vacuum,
is quite straightforward.

The situation is very different
in models with extended
Higgs sectors. For instance in the general two-Higgs-doublet model~(THDM),
introduced by T.D.~Lee~\cite{Lee:1973iz} in order to allow for CP violation in the Higgs sector,
we encounter already three bilinear terms as well as seven
quartic terms in the potential, corresponding to 14 real parameters.
For some recent works on the THDM we
refer to~\cite{Nishi:2007dv,Ginzburg:2009dp,Ferreira:2010yh,Maniatis:2011qu,Branco:2011iw}.
Much motivation for the
two-Higgs doublet model arises from supersymmetry, since
supersymmetric extensions of the SM require to have at least
two Higgs doublets -- besides the corresponding superpartners.
The minimal supersymmetric extension of the Standard Model~(MSSM) has
exactly two Higgs doublets and is therefore
a special THDM; for a review see~\cite{Martin:1997ns}.
However the MSSM has a number of drawbacks,
in particular the so-called $\mu$~parameter in the superpotential
has to be adjusted
by hand to the electroweak scale, that is, the electroweak scale
appears {\em not} via electroweak symmetry breaking.
This $\mu$~problem may be circumvented by a vacuum-expectation value
of an additional (complex) Higgs-boson singlet, leading to
the next-to-minimal supersymmetric standard model~(NMSSM); for
reviews see~\cite{Maniatis:2009re,Ellwanger:2009dp}.
Different supersymmetric models have been proposed which
have two Higgs-boson doublets and one or several
Higgs-boson singlets; an overview can be found in~\cite{Barger:2006dh}.\\

Other examples of extended Higgs sectors appear in
the study of additional discrete symmetries. Extending the Higgs sector
and assigning all fermions and scalars to certain
irreducible representations with respect to the discrete symmetry
(for instance the quaternion group with eight elements, $Q_8$),
the masses and mixing angles of the quarks and leptons may
be given by very few parameters; see for instance~\cite{Grimus:2003kq,Frigerio:2004jg}.\\

While dealing with an extended Higgs sector it is mandatory to
minimize the Higgs potential.
The determination of the global minimum is essential
to determine the masses of the fermions via
the Yukawa couplings and to ensure that
the electroweak symmetry breaking behavior follows
$\eweakgroup \to \emgroup$. However, finding the global minimum of a given non-linear multivariate function is a highly non-trivial task. There are a number of numerical methods, e.g., conjugate gradient, steepest descent, simulated annealing, genetic algorithm, etc. which are devised to find the global minimum but for one reason or another almost all of them are known to fail even for a relatively small system: the multivariate functions usually come with many minima and the above mentioned methods can be easily stuck to a local minimum instead of the global minimum. Hence, one can never be sure that the actual global minimum is found by these methods.

A more systematic approach to minimize the potential
is given by the solution of the stationarity equations.
Suppose the potential is bounded from below, then the
stationary points with the lowest
potential value are the global minima, that is, the vacua.
However, again, solving the stationary equations amounts to
finding the solutions of a system of non-linear multivariate equations which is equally hard.
However, the finding of all stationary points not only reveals the global minimum,
but gives some valuable insight to the structure of the potential: in this
way all maxima, saddle points, local minima besides the global minima are detected.

If the potential and hence the system of stationary equations have polynomial-like 
non-linearity (as opposed to involving transcendental functions etc.), then one viable 
option is to use a symbolic method called the Gr\"obner basis approach. 
The Gr\"obner basis approach was used to find all the stationary points and hence
the global minimum of extended Higgs potentials in~\cite{Maniatis:2006jd}.
There, for a special case with two doublets and two real singlets, it was shown that
extended Higgs potentials may develop a very rich
structure of stationary points.

The Gr\"obner basis approach, however, has several severe drawbacks:
firstly, the Buchberger algorithm is known to suffer from the exponential
space complexity, i.e., the memory (RAM) required by the machine exponentially
blows up with increasing number of variables, number of equations,
degree of equations or number of monomials.
The Buchberger algorithm 
is highly sequential and hence very difficult to parallelize. 
Thus, practically, the Gr\"obner basis approach
is limited to only small polynomial systems of equations of
lower degrees. 

In particular, in the example of two doublets and $n$ real singlets it is currently
not possible to go far beyond the case~$n=2$, as considered in~\cite{Maniatis:2006jd}.

In this work we shall introduce the 
numerical polynomial homotopy continuation method~\cite{SW:95,Li:2003} as
a method to find all stationary points of extended Higgs potentials.
We shall demonstrate
that this method is very powerful and overcomes all of the above mentioned shortcomings of the Gr\"obner basis approach. The approach allows to identify
{\em all} stationary solutions and supposed the potential is bounded from
below the global minima are therefore certainly detected. As an example we
minimize the most general potential with two complex Higgs-boson doublets
and $n$ real Higgs-boson singlets. We do a detailed analysis for the cases up
to $n=5$ in this paper, but we emphasis that we can straightforwardly go up to
$n=10$ even on a regular desktop machine.
And by exploiting the parallelizable nature of the method,
it is possible to go even beyond the $n=10$ case.

\section{The Higgs potential and the stationarity equations}
\label{s:potential}
In this Section, besides fixing the  notation
we explicitly write down the potential and its stationary equations
along with the tadpole conditions.

Let us consider a Higgs potential with two Higgs-boson doublets
and $n$~Higgs-boson singlets~\cite{Maniatis:2006jd}.
By convention we assume that both doublets carry
hypercharge $y=+1/2$, and denote the
complex doublet fields by
\begin{equation}
\label{eq-doubldef}
\varphi_i(x) = \begin{pmatrix} \varphi^+_i(x) \\  \varphi^0_i(x) \end{pmatrix},
\quad
i=1,2 .
\end{equation}
For the singlets we assume real fields which we denote by
\begin{equation}
\phi_i(x), \quad i=1,\ldots,n .
\end{equation}
Of course any complex singlet field may be decomposed into two
real singlet fields.

We employ the bilinear approach for the
Higgs doublets (see~\cite{Maniatis:2006fs} for details).
The Higgs-boson doublets can only appear in the form of
scalar products
$\varphi_i^\dagger \varphi_j$ with $i,j \in \{1,2\}$ in the potential,
ensuring gauge invariance.
As shown in~\cite{Maniatis:2006fs} we may replace
the scalar products by the bilinears $K_0$, $K_1$, $K_2$
and $K_3$,
\begin{equation}
\label{eq-phik}
\varphi_1^{\dagger}\varphi_1 = (K_0 + K_3)/2, \quad
\varphi_1^{\dagger}\varphi_2 = (K_1 + i K_2)/2, \quad
\varphi_2^{\dagger}\varphi_2 = (K_0 - K_3)/2, \quad
\varphi_2^{\dagger}\varphi_1 = (K_1 - i K_2)/2.
\end{equation}
The bilinears have to fulfill the conditions
\begin{equation}
\label{eq-kconditions}
K_0 \ge 0, \quad K_0^2-K_1^2-K_2^2-K_3^2 \ge 0.
\end{equation}
As shown in~\cite{Maniatis:2006fs} the four quantities
$K_\alpha$, satisfying~\eqref{eq-kconditions}, parameterize the
gauge orbits of the Higgs doublets.
The advantage of the replacement of the Higgs-boson doublets
by the bilinears is that we eliminate the
\eweakgroup gauge degrees of freedom and moreover simplify
the potential since the $K_\alpha$ are bilinear in the
Higgs-boson doublets.
We thus end up with a potential of the form
\mbox{$V(K_0, K_1, K_2, K_3, \phi_1,\ldots,\phi_n)$}.

\subsection{Stationary points}
\label{stationarity}

The most general Higgs potential of
two Higgs doublets and~$n$ real singlets reads
\begin{equation}
\label{potential}
\begin{split}
V(K_0, K_1, K_2, K_3, \phi_1,\ldots,\phi_n)=
&
\xi_\alpha\, K_\alpha + \eta_{\alpha \beta}\, K_\alpha K_\beta +\\
&
a_i\, \phi_i +
b_{i j}\, \phi_i \phi_j +
c_{i j k}\, \phi_i \phi_j \phi_k +
d_{i j k l}\, \phi_i \phi_j \phi_k \phi_l +\\
&
e_{i\, \alpha}\, \phi_i K_\alpha +
f_{i j\, \alpha}\, \phi_i \phi_j  K_\alpha
\end{split}
\end{equation}
with $\alpha, \beta \in \{0,...,3\}$ and $i,j,k,l \in \{1,...,n\}$
and summation convention for repeated indices.
Any additional term violates \eweakgroup invariance
respectively renormalizability. Note that the coefficients
$\eta_{\alpha \beta}$, $b_{ij}$, $c_{ijk}$, $d_{ijkl}$, and $f_{ij\,\alpha}$ are
symmetric in their greek respectively latin indices.
Depending on the number of real singlets~$n$ we count
$
14 + 5 n +
5 \binom{n+1}{2} +
\binom{n+2}{3} +
\binom{n+3}{4}
$ coefficients in the most general case. Hence
for the case $n=5$ we have 219 coefficients in the
general case.

In order to determine the stationary points of the Higgs
potential~\eqref{potential}
we take the constraint \eqref{eq-kconditions} into account.
We distinguish the possible cases of stationary points
with respect to the \eweakgroup symmetry breaking behaviour~\cite{Maniatis:2006fs,Maniatis:2006jd}:
\begin{itemize}\label{eq-kcondnon}
\item{{Unbroken electroweak gauge symmetry}: This is a stationary point with
\begin{equation}
K_0=K_1=K_2=K_3=0.
\end{equation}
A global minimum of this type
has vanishing vacuum expectation
values for the doublet fields~(\ref{eq-doubldef}).
The stationary points of this type are found for vanishing
Higgs-doublet fields, corresponding to vanishing bilinears $K_\alpha$
and requiring a vanishing
gradient with respect to the remaining real singlets:
\begin{equation}\label{eq-stationarityU}
\nabla \; V(\phi_1, \ldots, \phi_n) =0.
\end{equation}
}
\item{{Fully broken electroweak gauge symmetry}: This is a stationary point with
\begin{equation}\label{eq-kcondfull}
K_0>0,\qquad K_0^2-K_1^2-K_2^2-K_3^2 > 0.
\end{equation}
A global minimum of this type has non-vanishing
vacuum-expectation values for the
charged components of the doublets fields in~(\ref{eq-doubldef}),
thus gives fully broken~\eweakgroup; see~\cite{Maniatis:2006fs}.
The stationary points of this type are found by requiring a vanishing gradient
with respect to all bilinears $K_\alpha$ and all singlet fields:
\begin{equation}
\label{eq-stationarityF}
\nabla \; V(K_0, K_1, K_2, K_3, \phi_1, \ldots, \phi_n) =0.
\end{equation}
In this case the constraints~\eqref{eq-kcondfull} on the bilinears
must be checked explicitly for the real solutions.
}
\item{{Partially broken electroweak gauge symmetry}: This is a stationary point with
\begin{equation}\label{eq-kcondpartial}
K_0>0,\qquad
K_0^2-K_1^2-K_2^2-K_3^2 = 0.
\end{equation}
For a global minimum of this type follows the desired partial breaking
of~\eweakgroup down to~\emgroup; see~\cite{Maniatis:2006fs}.
Using the Lagrange multiplier method, these stationary points are given
by the real solutions of the system of equations
\begin{equation}\label{eq-stationarityP}
\begin{split}
\nabla
\big[
V(K_0, K_1, K_2, K_3, \phi_1, \ldots, \phi_n)
       - u \cdot (K_0^2-K_1^2-K_2^2-K_3^2)\; \big] &= 0,
\\
K_0^2-K_1^2-K_2^2-K_3^2 &= 0,
\end{split}
\end{equation}
where $u$ is a Lagrange multiplier.
The inequality in~\eqref{eq-kcondpartial} must be
checked explicitly for the real solutions found for~\eqref{eq-stationarityP}.
}
\end{itemize}
If the potential is bounded from below,
the global minima will be among the stationary solutions.
Plugging the solutions into the potential
the global minima are
the solutions with the lowest potential value.

The systems of stationary equations,
\eqref{eq-stationarityU},
\eqref{eq-stationarityF}, \eqref{eq-stationarityP}
are non-linear, multivariate, inhomogeneous systems
of polynomial equations
of degree $3$. For the case of two doublets and $n$ real singlets we have for
the unbroken, fully broken, and the partially broken
systems sets with $n$, $4+n$, and $5+n$
equations, respectively.
The number of indeterminants equals
the number of equations.

\section{Numerical polynomial homotopy continuation method}
\label{s:homotopy}
Finding the global minima or all
stationary points of a given potential
is usually a very difficult problem. As outlined
in the Introduction, a systematic approach is
to solve the stationarity systems of equations.
While the problem
of finding all solutions which include minima,
maxima and saddle points, compared to only finding minima,
seems more complicated at first sight, a powerful
method called the Gr\"obner basis method
can be used to solve the stationary equations as
long as they are polynomial equations.
However, due to the algorithmic complexity issues discussed in the Introduction,
we are practically restricted to solve only small systems.
In particular, going beyond the NMSSM, involving
two Higgs-boson doublets and one complex singlet (equivalent to two real singlets),
treated in~\cite{Maniatis:2006jd},
appears to be difficult to handle
with the Gr\"obner basis approach.

Here, we present a novel numerical method, the numerical
polynomial homotopy continuation (NPHC) method, which overcomes
all the shortcomings of the Gr\"obner basis method.
The NPHC method was recently introduced in particle theory
and statistical mechanics areas in \cite{Mehta:2009} to find all the so-called Gribov copies of the Landau gauge-fixing conditions on the lattice \cite{vonSmekal:2007ns,vonSmekal:2008es,Mehta:2010pe} and
subsequently found many applications in many areas of
theoretical physics including string phenomenology, lattice field theories, theoretical chemistry, non-linear dynamics, 
etc.~\cite{Mehta:2009zv,Mehta:2011xs,Mehta:2011wj,KastnerMehta11,2012arXiv1202.3320M}.

The basic strategy behind numerical continuation methods is:
first find the solutions of a simple system of equations
which shares several important features with the given system.
Then, in a second step, starting from each of these solutions
one continues them towards the given system in a systematic way.
Homotopy continuation methods have been around already
for several decades~\cite{Roth:62,AllgowerGeorg}, but with more
recent machinery like the NPHC method used in the present article,
the method is guaranteed to find all isolated solutions of systems
of polynomial equations \cite{SW:95,Li:2003}.

Suppose we want to solve a system of $m$ polynomial equations
\begin{equation}\label{e:poly}
P(x)=\begin{pmatrix}p_{1}(x)\\\vdots\\p_{m}(x)\end{pmatrix}=0
\end{equation}
in the variables $x=(x_{1},\dots,x_{m})^{T}$, which is known to
have only isolated solutions. Then the B\'ezout's Theorem
asserts that a system of $m$ polynomial equations in~$m$
variables has at most $\prod_{i=1}^m d_i$ isolated complex (which obviously include real)
solutions where $d_i$ is the degree of the $i$th polynomial.
This upper bound is called the \emph{classical B\'ezout bound},
and it is known to be sharp for generic systems
(i.e., for generic values of the coefficients of the polynomials $p_i(x)$).

Then, we formulate a homotopy (or, a set of parametric equations) as
\begin{equation}\label{e:homo}
H(x,t) = P(x) (1-t) + \gamma\, t\, Q(x),
\end{equation}
where $\gamma$ is a complex number and
\begin{equation}\label{e:polystart}
Q(x)=\begin{pmatrix}q_{1}(x)\\\vdots\\q_{m}(x)\end{pmatrix}=0
\end{equation}
is again a system of $m$ polynomial equations. Now, varying
the parameter $t\in[0,1]$, $H$ can be deformed from
the {\em start system} $H(x,1)=\gamma\, Q(x)$ at $t=1$ into
the polynomial system we want to solve, $H(x,0)= P(x)$ at $t=0$.
The following conditions have to be satisfied in order to guarantee
that all solutions of $P$ can be computed from this homotopy:
\begin{enumerate}
\item The solutions of $Q(x)=0$ can be computed relatively straightforwardly.
\item The number of solutions of $Q(x)=0$ satisfies the classical B\'ezout bound for $P(x)=0$ as an equality.
\item The solution set of $H(x,t)=0$ for $t\in(0,1]$ consists of a finite number of smooth paths, called {\em homotopy paths}, which are parameterized by $t$.
\item Every isolated solution of $H(x,0)=P(x)=0$, including the multiple roots, can be reached by some path originating at a solution of $H(x,1)=\gamma Q(x)=0$.
\end{enumerate}
Satisfying the first two criteria hinges on a suitable choice
of the start system $Q$. Criteria 3 and 4 are guaranteed to be satisfied,
for generic constants $\gamma$ in \eqref{e:homo}~\cite{SW:95}.

The start system $Q(x)=0$ can, for example, be taken to be
\begin{equation}\label{eq:Total_Degree_Homotopy}
Q(x)=\begin{pmatrix}
x_{1}^{d_{1}}-1\\
\vdots\\
x_{m}^{d_{m}}-1
\end{pmatrix}=0,
\end{equation}
where $d_{i}$ is the degree of the $i^{th}$ polynomial of $P(x)=0$.
The system \eqref{eq:Total_Degree_Homotopy} is easy to solve and
guarantees that the total number of start solutions is $\prod_{i=1}^{m}d_{i}$
and all solutions are nonsingular.

Each homotopy path, starting at a solution of $Q(x)=0$ at $t = 1$, is
tracked to $t = 0$ using a path tracking algorithm, e.\,g., Euler predictor
and Newton corrector methods. There are a number of freeware packages
well-equipped with path trackers such as
PHCpack~\cite{Ver:99}, HOM4PS2~\cite{Li:03}, and Bertini~\cite{BHSW06}.
We have used Bertini and HOM4PS2 to obtain the results in this paper.
Tracking the solutions to $t=0$, the set of endpoints of these homotopy
paths is the set of all solutions to $P(x)=0$. Since each homotopy
path can be tracked independently, NPHC is inherently parallelizable.
The NPHC method does not suffer from any major computational complexity.
It takes floating point coefficients in very naturally.

The set of real solutions can be obtained from the set of complex
solutions by considering the imaginary part of the solutions
(typically, up to a numerical tolerance). We remark that the
approach of \cite{HS:12} implemented in alphaCertified~\cite{alphaCertified}
can be used to certify the reality or non-reality of a nonsingular
solution given a numerical approximation of the solution.
The ability to compute all complex solutions, and thus all
real solutions, distinguishes the NPHC method from most other methods.

\section{Minimization of the Higgs potential}
\label{s:minimization}

\begin{figure*}[ht]
\begin{center}
\includegraphics[width=0.9\linewidth]{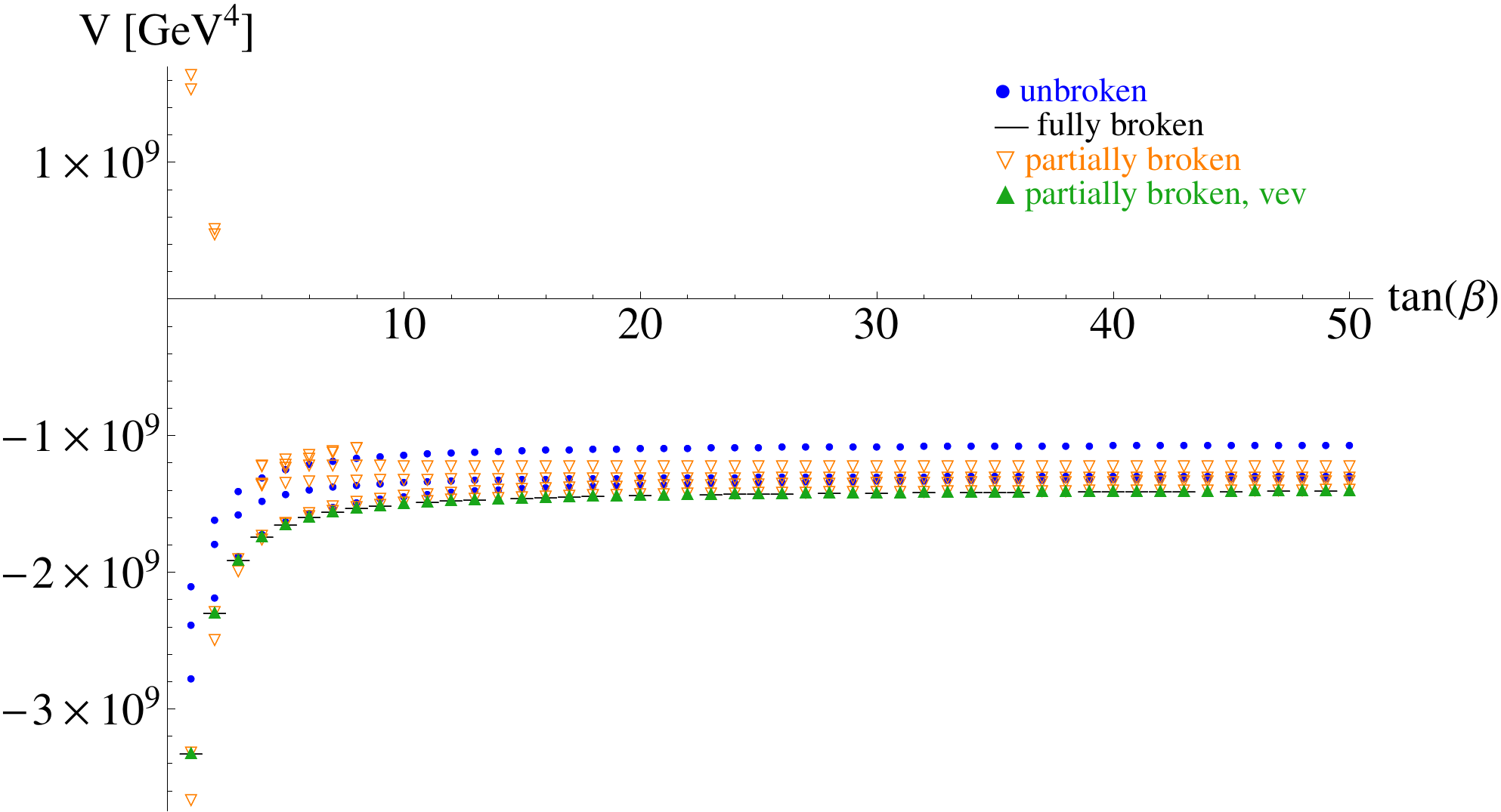}
\end{center}
\caption{\label{plotN2}
Higgs potential values for the stationary points
with 2 Higgs-boson doublets and
2 Higgs-boson singlets.
The parameter $\tan(\beta)$ is varied in the range $\tan(\beta)=1,\ldots, 50$.
All other parameters are fixed as described in the text.
The different types of stationary points are shown with
respect to the electroweak symmetry breaking behavior.
The filled dots denote the unbroken, the empty squares
the fully broken solutions and the triangles the partially
broken solutions, respectively. Only the full green triangles correspond
to stationary points with vevs as given in~\eqref{vevsnum}.
The stationary point with the lowest potential value
is the global minimum.}
\end{figure*}
\begin{figure*}[ht]
\begin{center}
\includegraphics[width=0.9\linewidth]{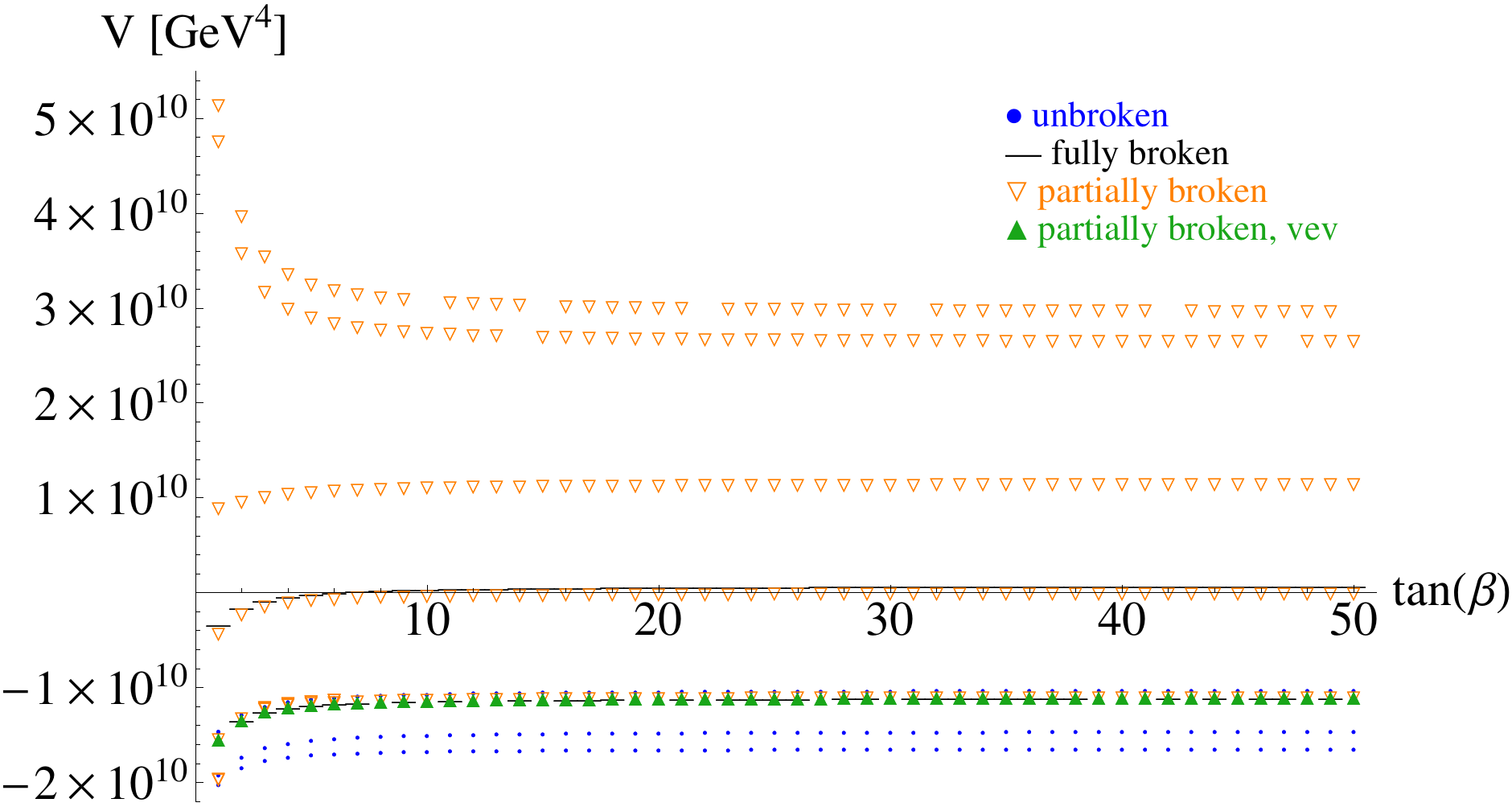}
\end{center}
\caption{\label{plotN3}
Similar to Fig.~\ref{plotN2} the Higgs potential values for the stationary points
with two Higgs-boson doublets but with 3 Higgs-boson singlets.
}
\end{figure*}
\begin{figure*}[ht]
\begin{center}
\includegraphics[width=0.9\linewidth]{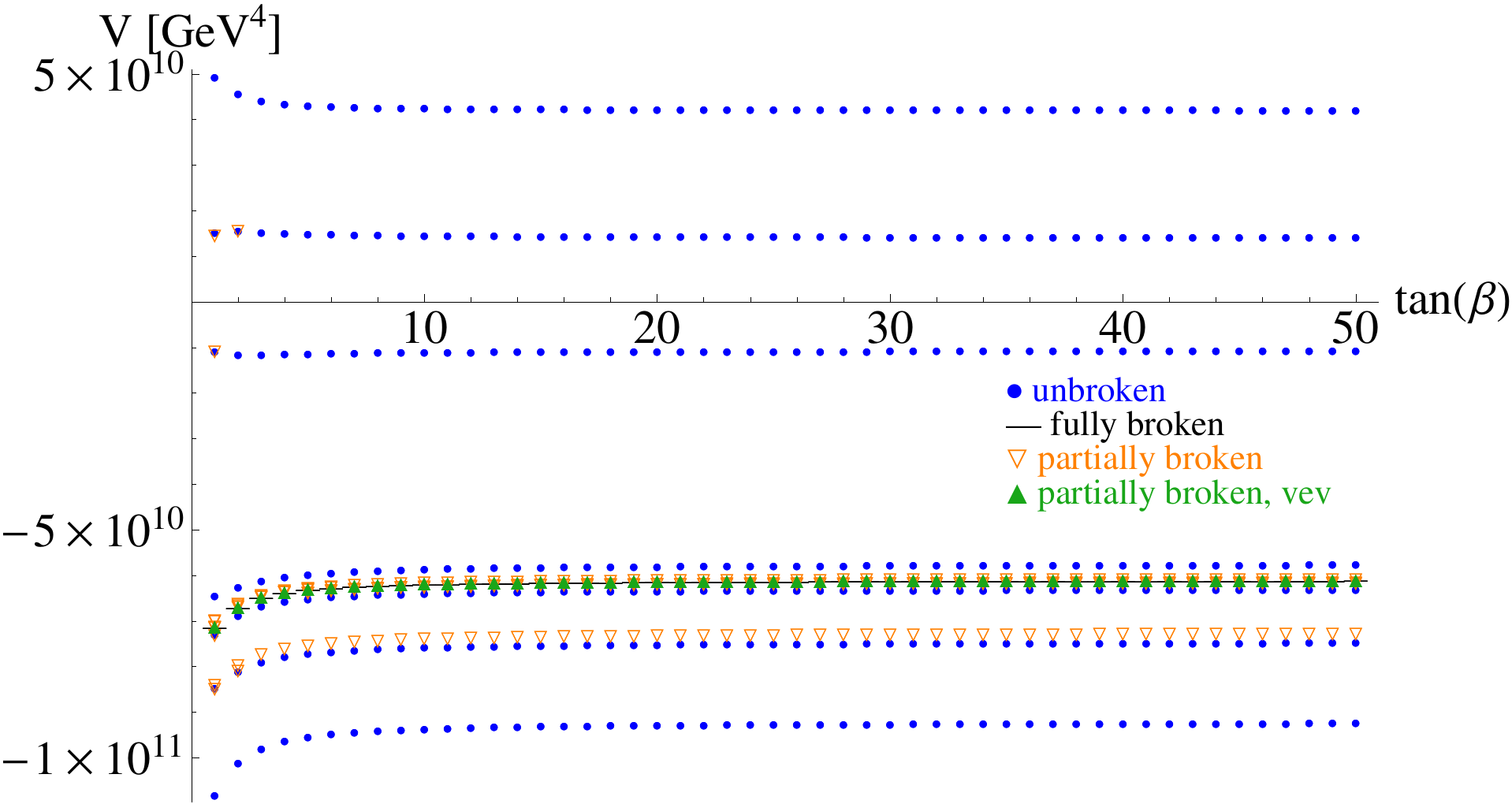}
\end{center}
\caption{\label{plotN4}
Similar to Fig.~\ref{plotN2} the Higgs potential values for the stationary points
with two Higgs-boson doublets but with 4 Higgs-boson singlets.
}
\end{figure*}
\begin{figure*}[ht]
\begin{center}
\includegraphics[width=0.9\linewidth]{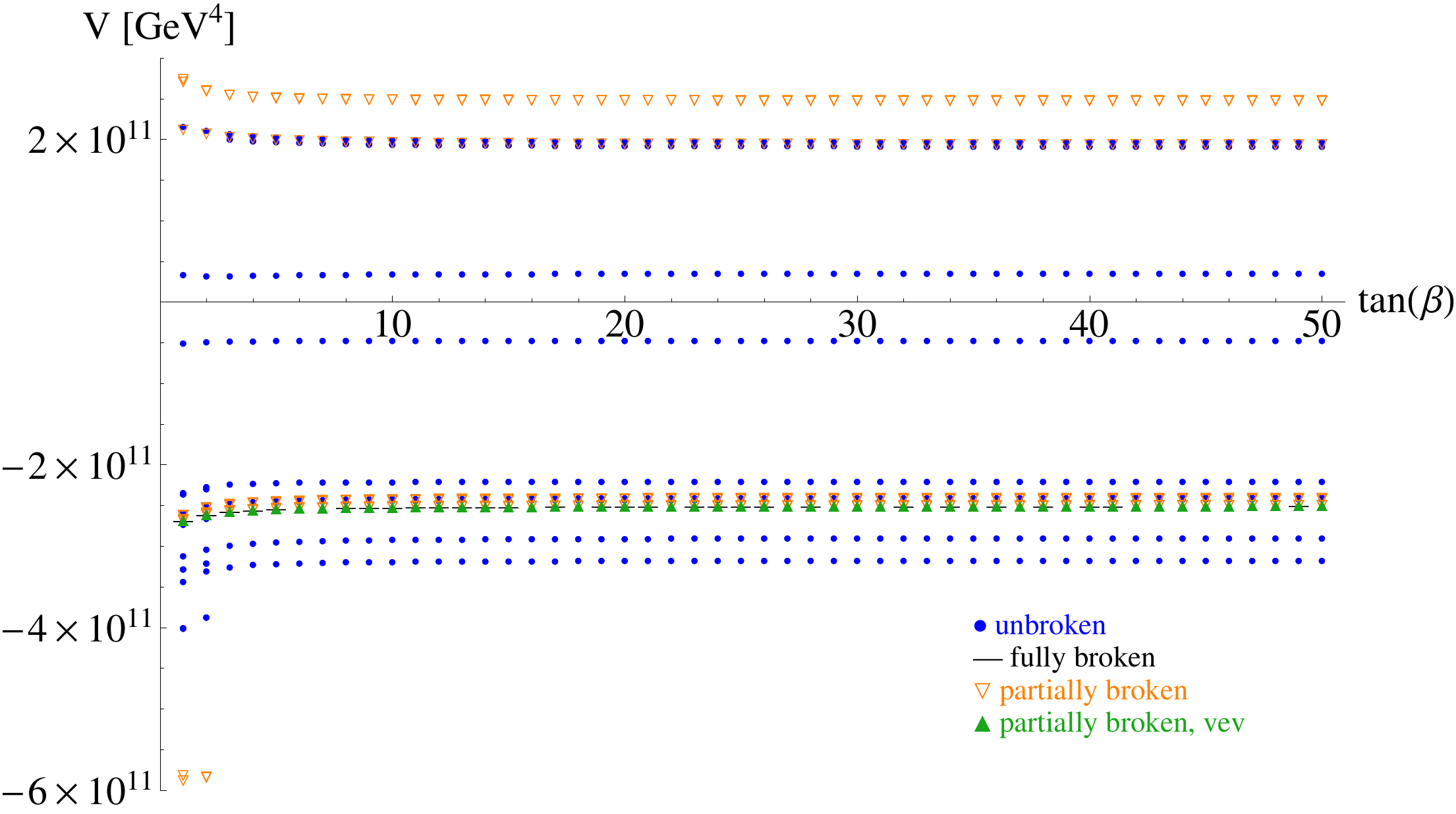}
\end{center}
\caption{\label{plotN5}
Similar to Fig.~\ref{plotN2} the Higgs potential values for the stationary points
with two Higgs-boson doublets but with 5 Higgs-boson singlets.
}
\end{figure*}

We now want to apply the NPHC method to the minimization problem
of the Higgs potential~\eqref{potential}, i.e., we want
to solve the stationarity systems of equations~\eqref{eq-stationarityU},
\eqref{eq-stationarityF}, and \eqref{eq-stationarityP}.
We note that the solution
with the correct electroweak
symmetry breaking has to come from the set~\eqref{eq-stationarityP}.
We start with numerically fixing the vacuum-expectation
values~(vevs) $v$, $\tan(\beta)$ and
$v'_i$, with $i=1,...,n$ with $n$ being the number of real singlets.
As usual $\tan(\beta)=v_2/v_1$ denotes the ratio of the vevs
of the two Higgs-boson doublets, $v^2 = v_1^2+v_2^2$
and $v'_i$ are the vevs of the real singlets:
\begin{equation} \label{vevs}
\langle \varphi_1 \rangle =
\frac{1}{\sqrt{2}} \begin{pmatrix} 0 \\ v_1 \end{pmatrix}, \qquad
\langle \varphi_2 \rangle =
\frac{1}{\sqrt{2}} e^{i\theta} \begin{pmatrix} 0 \\ v_2 \end{pmatrix}, \qquad
\langle \phi_i \rangle =
\frac{1}{\sqrt{2}} v'_i
\end{equation}
Except for~$\xi_\alpha$, $\alpha=0,...,3$ and $a_i$, $i=1,...,n$, we first fix
all coefficients in~\eqref{potential} numerically.
The parameters $\xi_\alpha$ and $a_i$ are then fixed
by employing the tadpole conditions,
\begin{equation} \label{tadpole}
\big\langle \nabla V(K_0, K_1, K_2, K_3, \phi_1, \ldots, \phi_n) \big\rangle =0 .
\end{equation}
In this way we gain solutions corresponding
to the correct electroweak symmetry breaking behavior
with the desired vevs.\\

For the case considered here with two Higgs-boson doublets and
up to five real singlets
we choose the vevs
\begin{equation} \label{vevsnum}
v=246~\text{GeV},\quad
v_1'=100~\text{GeV},\quad
v_2'=150~\text{GeV},\quad
v_3'=200~\text{GeV},\quad
v_4'=250~\text{GeV},\quad
v_5'=300~\text{GeV},
\end{equation}
a vanishing phase~$\theta$ of the second doublet
and vary the ratio of the doublet vevs
in the range~$\tan(\beta)=1, \ldots, 50$.
All coefficients in~\eqref{potential}, except for $\xi_\alpha$
and $a_i$, that is, $\eta_{00}$, $\eta_{01}, \ldots, f_{nn\,3}$
are fixed in the order they appear in the potential.
As a numerical example we assign the values
$0.1$, $0.2$, $0.3$, $0.4$, $0.5$, $0.1,\ldots, 0.5$, $0.1\ldots$ in turn.
Note the symmetry of the coefficients, that is, only
ordered indices appear, for instance $b_{21}$ is replaced
by $b_{12}$ in the implicit summation in~\eqref{potential} before
the values are assigned.

With all parameters numerically fixed we solve the
systems of equations~\eqref{eq-stationarityU},
\eqref{eq-stationarityF}, and \eqref{eq-stationarityP}.
As discussed in Sec.~\ref{s:homotopy} we employ the
NPHC method for the solution of the stationary equations.
We note that on a regular desktop machine we could solve the systems
for $n=10$ in around $16$ hours for a fixed set of parameters.
However, since we want to scan over a large range of parameter values,
due to the available limited computer resources, we restrict ourselves
to smaller systems, i.e., $n=2,\ldots,5$. The largest of these systems
can be solved in around $4$ hours for a given set of parameters on the desktop machine.

For two Higgs-boson doublets and $n=2,\ldots,5$ real singlets we
show in Table \ref{tablesol}
the number of complex stationary solutions
with respect to the fully, unbroken and partially broken cases.
\begin{table*}
\begin{tabular}{cccc}
n & unbroken & fully broken & partially broken\\
\hline
2 & 9 & 9 & 54\\
3 & 27 & 27 & 162\\
4 & 75 & 81 & 486\\
5 & 225 & 243 & 1458
\end{tabular}
\caption{\label{tablesol} Number of complex solutions found for
the most general Higgs potential with two doublets
and~$n$ singlets. The number of solutions are
given with respect to the systems of equations
\eqref{eq-stationarityU} (unbroken case),
\eqref{eq-stationarityF} (fully broken case),
and \eqref{eq-stationarityP} (partially broken case), respectively.
}
\end{table*}

From all complex solutions the real solutions are sorted out
(practically we require all imaginary parts to be smaller than $10^{-8}$).
The number of real solutions is typically much smaller and depends on
the chosen parameters. The fact that the number of complex
solutions is constant with varied parameters is a good
indication that no solutions are missed.
For the fully broken set subsequently the conditions
\eqref{eq-kcondfull} are checked. Similarly,
for the partially broken case, the solutions
have to fulfill~\eqref{eq-kcondpartial}.

The potential values for all stationary solutions passing
all checks are plotted in Figs.~\ref{plotN2}, \ref{plotN3},
\ref{plotN4}, \ref{plotN5}
for 2, 3, 4, 5 real singlets, respectively.
The parameter~$\tan(\beta)$ is varied in the range
$\tan(\beta)=1,\ldots, 50$.
Since the potential is bounded from below the
stationary solutions with the lowest potential value
are the global minima. The stationary solutions
are denoted by circles, squares, and triangles corresponding
to unbroken, fully broken, and partially broken solutions, respectively.
In case of the partially broken solutions empty triangles
show solutions which do not give the required vevs~\eqref{vevsnum},
whereas the full triangles show the solutions which yield
to the required vevs (with a precision of 8~digits).

In order to check whether the partially broken solutions
give the desired vevs~\eqref{vevsnum} we calculate the vevs
from the solutions via~\eqref{vevs},
that is,
\begin{equation} \label{vevcheck}
\sqrt{2 K_0}= v, \qquad
\sqrt{\frac{K_0-K_3}{K_0+K_3}} = \tan(\beta), \quad
\sqrt{2} \; \phi_i = v_i'.
\end{equation}

We find a very rich structure of stationary points.
Obviously the global minimum, that is the vacuum, not
always correspond to the physically desired solution.
As can be seen in Fig.~\ref{plotN2} in the case
of two real singlets, the global minimum
breaks electroweak gauge symmetry partially,
however yields undesired vevs for
small $\tan(\beta)$ values.

Let us note that quite generically we find
nearly degenerate potential values for different
stationary solutions. 
Note that even when the potential values are nearly degenerate the
solutions may be located at very different field values.
For instance in the case of two Higgs-boson
doublets and two singlets we find for $\tan(\beta)=10$
the two deepest stationary solutions as shown explicitly
in Tab.~\ref{table-exp}.
\begin{table}
\begin{center}
\begin{tabular}{ccccccccc}
 & $K_0$ & $K_1$ & $K_2$ & $K_3$ & $\phi_1$ & $\phi_2$ & $u$ & $V$\\
\hline
solution 1:&
30257&
5991.6&
0.0046390&
-29658&
70.710&
106.06&
$-1.0334\cdot 10^{-8}$&
$-1.5045\cdot 10^9$\\
solution 2:&
34878&
8085.9&
-4701.0&
-33600&
55.630&
123.54&
-0.0036596&
$-1.5051\cdot10^9$
\end{tabular}
\end{center}
\caption{\label{table-exp}The two deepest solutions explicitly for the
case of two singlets and $\tan(\beta)=10$. Solution 1 corresponds
to the desired vevs, whereas solution 2 gives wrong vevs.}
\end{table}
In this numerical example only the solution~1 corresponds to the
desired vevs~\eqref{vevsnum}, however solution~2 is the global minimum.
This behavior of nearly degenerate vacua
was already mentioned for the case of the NMSSM in~\cite{Maniatis:2006jd}
which we confirm for more general cases.

\clearpage

\section{Conclusions}
\label{s:conclusions}

The extended Higgs models have attracted a lot of attention recently,
in particular with the on-going experiments at the Large Hadron Collider.
The main technical difficulty in studying these models is that
accurately finding the global minimum of the corresponding Higgs
potential is highly non-trivial due to its non-linear nature.
Whereas the Higgs potential of the Standard Model
is rather simple, this is in general not longer true for extended potentials.
The solution of the stationarity equations allows to identify
the vacuum, as long as the potential is bounded from below.
In this paper, we have applied the numerical polynomial homotopy
continuation method to solve 
the non-linear, multivariate systems of polynomial stationary equations
for rather involved systems of Higgs potentials.
In contrast to most minimization methods this method
guarantees the detection of the global minimum given by
the stationary solution with the lowest potential value.

We have shown that the case of the most general
Higgs potential with two complex doublets and five real
singlets is solvable via numerical polynomial homotopy continuation.
The results show a rich structure
of stationary solutions with local maxima, saddle points and minima.
In this way,
we have shown that a large set of parameter values can be
excluded from the physically viable regions.
Among the solutions
we typically find solutions with nearly degenerate
potential values, however far away in terms of
the Higgs-boson fields. Let us note
that the numerical polynomial homotopy continuation method
allows to go gar beyond the limits of two doublets and five singlets.

\begin{acknowledgments}
D.M. acknowledges support by the U.S. Department of Energy under contract DE-FG02-85ER40237.
\end{acknowledgments}

\bibliographystyle{h-physrev}
\addcontentsline{toc}{section}{\refname}\bibliography{bibliography_homotopy}

\end{document}